# Hybrid electromagnetic toroidal vortices

Ren Wang[1,2*], Beier Ying[1], Shuai Shi[1], Junsong Wang[1], Bing-Zhong Wang[1], Musheng Liang[1,2*], and Yijie Shen[3,4*]

[1] *Institute of Applied Physics, University of Electronic Science and Technology of China, Chengdu 611731, China*

[2] *Yangtze Delta Region Institute (Huzhou), University of Electronic Science and Technology of China, Huzhou 313098, China*

[3] *Centre for Disruptive Photonic Technologies, School of Physical and Mathematical Sciences, Nanyang Technological University, Singapore 637378, Singapore*

[4] *School of Electrical and Electronic Engineering, Nanyang Technological University, Singapore 637378, Singapore*

* E-mail: rwang@uestc.edu.cn (R.W.); msliang@uestc.edu.cn (M.L.); yijie.shen@ntu.edu.sg (Y.S.)

## Abstract

**The ubiquitous occurrence of toroidal vortices or vortex rings in fluid-dynamic scenarios in nature has garnered significant attention of scientific frontier, whilst, the electromagnetic counterparts of which were only proposed recently with two distinct manifestations: vector toroidal pulses [Nat. Photon. 16, 523 (2022)] and scalar phase toroidal vortices [Nat. Photon. 16, 519 (2022)]. This dichotomy in the understanding of toroidal vortex phenomena has prompted a reassessment of their fundamental nature. Herein, we theoretically propose a novel form of electromagnetic toroidal vortex solutions, that uniquely integrate both scalar and vector characteristics, challenging the prevailing notion of their mutual exclusivity. We also present the experimental generation of the hybrid toroidal vortex pulses by a compact coaxial horn emitter augmented with a metasurface. This methodology not only demonstrates the feasibility of creating such complex vortex structures but also endows the resulting pulses with unique properties, including the coexistence of transverse orbital angular momentum, electromagnetic vortex streets, and topological skyrmion textures. These attributes introduce new dimensions in topologically complex structured waves, opening avenues for enhanced free-space information transmission, topologically nontrivial light-matter interaction and microscopy techniques.**

## Introduction

The phenomenon of toroidal vortices, also known as vortex rings, is widely observed in nature [1]. These vortices manifest in various forms such as bubble rings, smoke rings, and mushroom clouds, and have been seen in the locomotion of flagellates [2], spore dispersal [3], dandelion flight [4], cumulus clouds [5], drop splashing [6], and blood flow through the heart [7]. The distinctive and intriguing topological quasiparticle structure of toroidal vortices has led to extensive research across multiple scientific fields. These vortices have been identified or intentionally created in diverse environments, including tokamaks [8], nuclei [9], heterogeneous media [10], Bose-Einstein condensates [11], pseudospin structures [12], magnetic bubbles [13], droplet manipulation [14], and shock-accelerated interfaces [15]. In addition to isolated toroidal vortices, the interactions between different toroidal vortices have garnered significant scholarly interest [16]-[18]. Hybrid toroidal vortices have been observed in Bose-Einstein condensates [19],[20], magnetic nanocylinders [21], and generic excitable media [22].

In the realm of electromagnetics, scholars have proposed two kinds of independent electromagnetic toroidal vortices: vector and scalar [23]. Vector electromagnetic toroidal vortices, denoted as field-line toroidal pulses [24], are characterized by an array of neighboring electric or magnetic field lines, delineating a toroidal surface at a distance from the optical axis, as shown in Fig. 1(a). They exhibit skyrmion topology [25]-[27], possess anapole-excitation capability [28], and have been utilized for three-dimensional superresolution localization [29]. In contrast, scalar electromagnetic toroidal vortices, referred to as phase-vortex toroidal pulses [30], induce a phase variation of $2\pi$ around the surface of a torus along a poloidal coordinate, as shown in Fig. 1(b). This results in the formation of an optical vortex with torus structure, representing a higher-dimensional form of spatiotemporal vortices with transverse orbital angular momentum [31]-[35]. Recently, vector electromagnetic toroidal vortices have been observed through methodologies involving metasurfaces [24], quantum control [36], coaxial horn antennas [37], or gold nanotorus [38], whereas scalar electromagnetic toroidal vortices have been observed utilizing techniques like conformal transformations or symmetry-breaking gratings [30],[39],[40]. However, this dichotomy in the understanding of electromagnetic toroidal vortex phenomena has prompted a reassessment of their fundamental nature. It remains unknown whether scalar and vector toroidal vortices can coexist within a single electromagnetic pulse, making the observation of scalar and vector hybrid electromagnetic toroidal vortices (HETVs) still elusive.

In this paper, we proposed a new family of electromagnetic pulse that carry coupled scalar and vector HETVs and present a method for generating such pulses using coaxial horn antennas equipped with radial metasurfaces, as shown in Fig. 1. These metasurfaces enable the creation of scalar electromagnetic toroidal vortex. Importantly, we find the singularities of scalar toroidal vortex induces saddle points and further results in the generation of vector electromagnetic toroidal vortices. Consequently, we observed the scalar and vector HETVs with unique properties, including the coexistence of transverse skyrmion textures, electromagnetic vortex streets, and orbital angular momentum.

**Results**

**Generation scheme for HETVs.** The device for generating HETV in an electromagnetic pulse primarily consists of a coaxial horn emitter and specially designed radial metasurfaces, as shown in Fig. 1. The coaxial horn emitter, which features inner and outer conductors (Fig. 1(c)), has been proposed for generating electromagnetic vector toroidal pulses [37],[41]. The coaxial horn emitter can emit radially polarized electromagnetic waves, characterized by a radial polarization null point and a longitudinal polarization peak along the propagation axis [37]. To transform the radial polarization component into waves containing scalar toroidal vortices, we positioned a purposely designed radial metasurface in front of the coaxial horn emitter (Fig. 1(d1)). Considering an arbitrary radial range, transfer function of radial metasurface can be expressed as [42]:

$$H(k_r, \omega) \approx S_r k_r / k_0 + S_t (\omega - \omega_0) / \omega_0 \tag{1}$$

where $k_r$ is transverse wave vector along radial direction, $\omega$ is angular frequency of plane wave centered at $\omega_0$ with a wave vector $k_0$, $S_r$ and $S_t$ are two complex constants determining amplitude and phase distribution in $k_r$ - $\omega$ domain. This metasurface is coaxial with the emitter and composed of radially arranged units, each radius containing two asymmetric metasurface units, each made of a substrate and two oppositely oriented C-shaped copper foils (Fig. 1(d2)). Detailed structures and dimensions of these units are provided in the supplementary materials. Spatiotemporal vortex pulses can be generated based on the breaking of spatial mirror symmetry [18-21]. The two oppositely oriented C-shaped metallic slabs in the radial metasurfaces create an asymmetric structure, which produces a phase distribution of $e^{-il\theta}$ in the spatial frequency –frequency domain. Therefore, $S_r$ and $S_t$ can be normalized to 1 and a $\pi/2$ (orbital angular momentum (OAM) mode: $l = -1$) or $-\pi/2$ (OAM mode: $l = +1$) phase difference is required between $S_r$ and $S_t$. A radially polarized output field $E_r(r,t)$ can be

$$E_r(r,t) = s(r)h(r,t)\exp(-i\omega_0 t + ik_0 z) \tag{2}$$

where $s(r)$ is the radially polarized component of incident wave, $h(r,t)$ is the inverse Fourier transform of the transfer function of radial metasurfaces. The longitudinal components $E_z$ can be determined by applying Gauss's law in free space, followed by numerical integration [24]:

$$E_z(r,z) = -\int_\alpha^z \frac{E_r(r,z^{'})}{r} + \frac{\partial E_r(r,z^{'})}{\partial r} dz^{'} \qquad (3)$$

where $\alpha$ is selected as a reference point where the field is zero. Combining the longitudinally polarized component $E_z$ with the transversely polarized component $E_r$ enables the construction of the spatiotemporal vector field distribution, where the singularities of the scalar toroidal vortex and radial polarization induce saddle points. The inner saddle points of the vector toroidal vortices generated by this method are located along the propagation axis, while the outer saddle points are at the phase singularities of the scalar toroidal vortices. Thus, the vector and scalar toroidal vortices exhibit coupled topologies and spatial relationships, forming a new type of electromagnetic pulse: HETV, as shown in Fig. 1(e). Within a few half-wavelengths around the vortex singularity in the HETV, several vector toroidal vortices connect, forming an electromagnetic vortex street and several skyrmion textures, as detailed in the following results. Please see the supplementary materials for detailed derivation of the HETV.

Our asymmetric metasurface unit is designed with a periodic boundary, as detailed in the supplementary materials. As shown in Fig. 1(d1), the distances between subarray units vary along different radii in the radial metasurfaces. This necessitates the designed asymmetric metasurface unit to have periodic robustness. The asymmetric modulation of the incident pulse can be illustrated by the transmission spectrum function $T(k_x, \omega)$, where $\omega$ is the angular frequency of a plane wave, and $k_x$ is the wave vector component along the asymmetric metasurface unit. When varying the spacing period $p$ (corresponding to the spacing between subarray units in the radial metasurfaces), the phase and amplitude responses of the designed unit's transmission spectrum function are shown in Figs. 2(a1-a3) and 2(b1-b3), respectively. As the period $p$ increases, the vortex phase singularities shift gradually to higher frequencies. The positions of the singularities for different values of $p$ are shown in Fig. 2(c). When $p$ varies between 2.4 and 7.7 (corresponding to the distance range of the metallic structures in each radially arranged subarray), there is always one vortex phase singularity with the same handedness present in the 2-2.2 GHz range. This ensures that when incident pulses are in the 2-2.2 GHz range, the radial metasurfaces can convert radially polarized electromagnetic waves into spatiotemporal vortices at each radial position, thereby forming a scalar toroidal vortex. Due to asymmetric modulation and Fourier-transform properties, a phase singularity in the transmission spectrum function of the radial metasurface shown in Fig. 1(d1), located at point $(k_{r0}, \omega_0)$ in the $(k_r, \omega)$ domain, can be directly translated into the spacetime $(r, \tau)$ domain for the transmitted pulse.

**Scalar and vector toroidal vortex performances of generated HETVs.** We demonstrate the topologically protected generation of HETVs in simulations and experiments, showcasing scalar and vector toroidal vortices schematically in Figs. 3 and 4. The coaxial horn emitter and 144 radial metasurface subarrays are mounted on a foam support structure to create the HETV generator, with units along each radial direction starting from $r$ = 40 mm; details are provided in Methods and supplementary materials. The coaxial horn emitter is driven by a Gaussian signal at carrier frequency $\omega_0 = 2.1$ GHz in simulations, which slightly shifts to $\omega_0 = 2.2$ GHz in experiments due to material parameter and fabrication errors. The bandwidth of feeding signal is 0.2 GHz. HETV generation was experimentally observed in an anechoic chamber using a planar near-field measurement system, detailed in the supplementary materials.

Figs. 3(a1), (a2), and (a3) presents the theoretical, simulated, and measured 3D iso-intensity profile of the radial polarized electric field components of the HETV pulse, respectively. Semitransparency highlights the ring structure and reveals the hidden ring-shaped vortex core, with the outer layer in gray and the vortex core surface in brown for clarity. To depict the spiral phase rotating around the vortex core, four slices are taken radially (marked numerically as 1–4), and their phase profiles in local coordinates are shown in Figs. 3(b1), (b2), and (b3). All slices exhibit a $2\pi$ spiral phase, indicating the existence of orbital angular momentum. Within a local region around OAM center, OAM per photon at the chosen radial slices corresponding to the theoretical results is consistently $1.013\,\hbar$. In simulations, the OAM per photon at the chosen radial slices is $1.26\,\hbar$, $1.18\,\hbar$, $1.11\,\hbar$, and $1.23\,\hbar$, respectively, while in measurements, it is $1.28\,\hbar$, $1.20\,\hbar$, $1.22\,\hbar$, and $1.20\,\hbar$, respectively. Please see the supplementary materials for the OAM calculation method. The intensity and phase profiles confirm the pulse's nature as a scalar toroidal vortex. It is important to note that since this scalar toroidal vortex originates from a radially polarized vector wave, the spatiotemporal phase refers to radial directions at each slice. In contrast, for a linearly polarized scalar toroidal vortex, the spatiotemporal phase refers to a linear direction [30]. Figs. 3(c1), (c2), and (c3) display the theoretical, simulated, and measured 2D spatiotemporal electric field distribution at an $(r, t)$ plane, respectively. Due

to symmetry, only the positive $x$-axis region is shown, with similar field distributions on other radii detailed in the supplementary materials. A fork-shaped pattern around ($t = 0$, $x = 200$ mm) indicates a $2\pi$ vortex in the space-time domain, corresponding to the double-layer toroidal isosurfaces in Fig. 3(a1/a2/a3) and the observed phase vortex in Fig. 3(b1/b2/b3). Due to material and assembly errors, the measued results show slight differences in the positions of singularities at different slices compared to the simulated results. However, the overall scalar toroidal vortex topology is well-preserved, thereby validating the effectiveness of this method.

Since the toroidal parameter in scalar electromagnetic toroidal vortices is phase-based, the pulses with these vortices are narrowband, contrasting sharply with the ultrawideband pulses required to observe vector toroidal pulses [24],[36],[37]. In fact, when narrowband signals are fed into an isolated coaxial horn emitter without radial metasurfaces, the radiated wave's electric field fails to form a proper vector toroidal vortex near the propagation axis, as detailed in the supplementary materials. However, in the proposed design of a coaxial horn emitter with radial metasurfaces, the singularities of the scalar toroidal vortex and radial polarization can induce the formation of vector toroidal vortices, as illustrated in Fig. 4.

From Fig. 3(c1/c2/c3), it can be observed that at the positions of the fork-shaped pattern, the radial polarized electric field components of the pulses exhibit opposite signs around the vortex singularity, leading to the formation of saddle points at those locations and consequently generating vector toroidal vortices, as depicted in Fig. 4. Figs. 4(a1), (a2), and (a3) respectively illustrate the distribution of theoretical, simulated, and measured electric field vectors on the cross-section shown in Figs. 3(c1), (c2), and (c3), where at distant times from the vortex singularity ($t = 0$), the vector electric field maintains continuity off the propagation axis, with no saddle points outside the propagation axis. At and near the vortex singularity, within several half-wavelengths, saddle points emerge due to the radial polarization with reversed signs, as highlighted in the enlarged view: Figs. 4(b1), (b2), and (b3). Due to the nearly rotational symmetric structure of the coaxial horn with radial metasurfaces employed, similar vector field distributions are observed at different radii, detailed in the supplementary materials. Consequently, the spatiotemporal vortex singularities generated by radial metasurfaces induce HETVs, coupling scalar and vector toroidal vortices in topological structure and spatiotemporal positioning. Within several half-wavelengths around the vortex singularity, multiple vector toroidal vortices are interconnected, forming an electromagnetic vortex street, a phenomenon previously observed only in supertoroidal pulses [37]. As a contrast, when the coaxial horn without radial metasurfaces is fed by the same pulses with the same bandwidth, neither scalar nor vector toroidal vortices can be observed near the propagation axis, please see the supplementary materials for details.

Similar to vector toroidal vortices, HETVs also feature skyrmion topology, illustrated in Figs. 4(c1), (c2), and (c3). These skyrmionic textures vary across different transverse planes near the center of the electromagnetic vortex ring while maintaining a Néel-type helicity. The skyrmion number alternates between "±" on either side of the saddle points. Both experimental and simulated HETVs consistently exhibit a skyrmion number of approximately ±1, confirming the presence of skyrmionic textures. The comprehensive coverage of the field vector sphere [43] in theoretical, simulated, and experimental HETVs provides further validation of the skyrmion presence, as shown in Figs. 4(d1), (d2), and (d3), respectively, with detailed calculation methods outlined in the supplementary materials. The vector field results demonstrate that both the simulated and measured data align well with the theoretical predictions, clearly exhibiting an HETV topology with skyrmionic textures.

## Discussion

In conclusion, we propose a novel class of electromagnetic pulse solutions, known as HETVs. These pulses carry coupled scalar and vector toroidal vortices, and we have demonstrated a method for their generation using coaxial horn antennas equipped with radial metasurfaces. Our investigation into the performance of scalar and vector toroidal vortices in this proposed scheme confirms the successful generation of HETVs.

Building on this approach, there is potential to extend the concept to other types of HETVs. Specifically, the scheme can potentially generate transverse-electric microwave toroidal pulses using an azimuthally polarized coaxial horn antenna with circumferential metasurfaces. The realization of an azimuthally polarized coaxial horn could involve substituting the inner and outer conductors with artificial magnetic conductors. Furthermore, this approach can be expanded into the terahertz and optical frequency ranges. Toroidal pulses have already been generated in these frequency ranges [24],[36], and recent advancements in asymmetric metasurfaces or metagratings offer pathways to generate optical and acoustic spatiotemporal vortices [18-21]. Integrating these vector toroidal pulse generation devices with spatiotemporal scalar vortex generation surfaces can facilitate the production of HETVs in corresponding frequency bands.

HETVs present practical opportunities for novel interactions between electromagnetic waves and matter. The space-time skyrmion textures within free-space HETVs are of particular interest for information transfer, akin to the interest in localized skyrmions for data storage in topological materials [44]. The scalar toroidal vortex topology also positions HETVs as promising candidates for free-space information transfer. The coupling of scalar and vector transmission characteristics within HETVs introduces new dimensions to classic entanglement and new possibilities for information transfer. Additionally, the scalar and vector singularities within HETVs hold promise for achieving super-resolution sensing and imaging. In summary, the emergence of HETVs represents a significant step towards innovative applications in telecommunication, metrology, and microscopy systems.

**Methods**

**Fabrication and measurement method.** Each metasurface subarray with 2 units was fabricated utilizing the printed circuit board technique. The substrate employed has a thickness of 1.5 millimeters, accompanied by a relative dielectric constant of 16 and a loss tangent of 0.001. The coaxial horn emitter, whose configuration and dimensions are congruent with those outlined in [37]. To assemble the HETV generator, the coaxial horn emitter is mounted alongside 144 radial metasurface subarrays onto a foam holder. These units are positioned along each radial direction, commencing from a radius of 40 millimeters. We employed a planar microwave anechoic chamber for comprehensive measurements of the spatial electromagnetic fields emitting from the HETV generator. By interconnecting a vector network analyzer to both the HETV generator and the probes, we procured transmission response measurements, enabling us to discern the magnitude and phase signatures of the electromagnetic field across diverse spatial locales. To accurately capture both transverse and longitudinal electric field components, we utilized a rectangular waveguide probe and a monopole probe, respectively. For a detailed account of our methodology, please refer to the supplementary materials. During the transverse component measurements, we precisely adjusted the polarization direction of the rectangular waveguide probe to align with the radial axis of the HETV generator. Similarly, for the longitudinal component assessments, we aligned the polarization of the monopole probe with the axial direction of the generator. Our scanning system was programmatically orchestrated to traverse the designated plane, facilitating the acquisition of a comprehensive field distribution. By harnessing the frequency domain measurement method outlined, we extracted the spatial magnitude and phase characteristics of the HETV generator. Subsequently, we employed inverse Fourier transformation to reconstruct the time-domain field, please see the supplementary materials for details. The synthesis of the transverse and longitudinal electric field components within the spatiotemporal field distribution enabled us to construct a comprehensive scalar and vector field representations, offering profound insights into the electromagnetic behavior of the HETV generator.

**References**

1. Matsuzawa, T., Mitchell, N. P., Perrard, S., & Irvine, W. T. Creation of an isolated turbulent blob fed by vortex rings. *Nat. Phy.* **19**(8), 1193-1200 (2023).

2. Hess, S., Eme, L., Roger, A. J., & Simpson, A. G. A natural toroidal microswimmer with a rotary eukaryotic flagellum. *Nat. Microbiology* **4**(10), 1620-1626 (2019).

3. Whitaker, D. L., & Edwards, J. Sphagnum moss disperses spores with vortex rings. *Science* **329**(5990), 406-406 (2010).

4.  Cummins, C., Seale, M., Macente, A., Certini, D., Mastropaolo, E., Viola, I. M., & Nakayama, N. A separated vortex ring underlies the flight of the dandelion. *Nature* **562**(7727), 414-418 (2018).

5.  Eytan, E., Arieli, Y., Khain, A., Altaratz, O., Pinsky, M., Gavze, E., & Koren, I. The role of the toroidal vortex in cumulus clouds' entrainment and mixing. *Journal of Geophysical Research: Atmospheres* **129**(14), e2023JD039493 (2024).

6.  Lee, J. S., Park, S. J., Lee, J. H., Weon, B. M., Fezzaa, K., & Je, J. H. Origin and dynamics of vortex rings in drop splashing. *Nat. Commun.* **6**(1), 8187 (2015).

7.  Kilner, P. J., Yang, G. Z., Wilkes, A. J., Mohiaddin, R. H., Firmin, D. N., & Yacoub, M. H. Asymmetric redirection of flow through the heart. *Nature* **404**(6779), 759-761 (2000).

8.  Choi, G. J., & Hahm, T. S. Long term vortex flow evolution around a magnetic island in tokamaks. *Phys. Rev. Lett.* **128**(22), 225001 (2022).

9.  Nesterenko, V. O., Repko, A., Kvasil, J., & Reinhard, P. G. Individual Low-Energy Toroidal Dipole State in Mg 24. *Phys. Rev. Lett.* **120**(18), 182501 (2018).

10. Kartashov, Y. V., Malomed, B. A., Shnir, Y., & Torner, L. Twisted toroidal vortex solitons in inhomogeneous media with repulsive nonlinearity. *Phys. Rev. Lett.* **113**(26), 264101 (2014).

11. Ruostekoski, J., & Anglin, J. R. Creating vortex rings and three-dimensional skyrmions in Bose-Einstein condensates. *Phys. Rev. Lett.* **86**(18), 3934 (2001).

12. Lim, L. K., & Moessner, R. Pseudospin vortex ring with a nodal line in three dimensions. *Phys. Rev. Lett.* **118**(1), 016401 (2017).

13. Zhao, X., Quinto-Su, P. A., & Ohl, C. D. Dynamics of magnetic bubbles in acoustic and magnetic fields. *Phys. Rev. Lett.* **102**(2), 024501 (2009).

14. An, D., Warning, A., Yancey, K. G., Chang, C. T., Kern, V. R., Datta, A. K., ... & Ma, M. Mass production of shaped particles through vortex ring freezing. *Nat. Commun.* **7**(1), 12401 (2016).

15. Wadas, M. J., Khieu, L. H., Cearley, G. S., LeFevre, H. J., Kuranz, C. C., & Johnsen, E. Saturation of vortex rings ejected from shock-accelerated interfaces. *Phys. Rev. Lett.* **130**(19), 194001 (2023).

16. Aref, H., & Zawadzki, I. Linking of vortex rings. *Nature* **354**(6348), 50-53 (1991).

17. Chatelain, P., Kivotides, D., & Leonard, A. Reconnection of colliding vortex rings. *Phys. Rev. Lett.* **90**(5), 054501 (2003).

18. Lim, T. T., & Nickels, T. B. Instability and reconnection in the head-on collision of two vortex rings. *Nature* **357**(6375), 225-227 (1992).

19. Ginsberg, N. S., Brand, J., & Hau, L. V. Observation of hybrid soliton vortex-ring structures in Bose-Einstein condensates. *Phys. Rev. Lett.* **94**(4), 040403 (2005).

20. Komineas, S., & Brand, J. Collisions of solitons and vortex rings in cylindrical Bose-Einstein condensates. *Phys. Rev. Lett.* **95**(11), 110401 (2005).

21. Liu, Y., & Nagaosa, N. Current-Induced Creation of Topological Vortex Rings in a Magnetic Nanocylinder. *Phys. Rev. Lett.* **132**(12), 126701 (2024).

22. Winfree, A. T. Persistent tangled vortex rings in generic excitable media. *Nature* **371**(6494), 233-236 (1994).

23. Cardano, F., & Marrucci, L. Smoke rings of light. *Nat. Photon.* **16**(7), 476-477 (2022).

24. Zdagkas, A., McDonnell, C., Deng, J., et al. Observation of toroidal pulses of light. *Nat. Photon.* **16**, 523–528 (2022).

25. Shen, Y., Hou, Y., Papasimakis, N., & Zheludev, N. I. Supertoroidal light pulses as electromagnetic skyrmions propagating in free space. *Nat. Commun.* **12**(1), 5891 (2021).

26. Shen, Y., Zhang, Q., Shi, P., Du, L., Yuan, X., & Zayats, A. V.. Optical skyrmions and other topological quasiparticles of light. *Nat. Photon.* **18**(1), 15-25 (2024).

27. Shen, Y., Papasimakis, N., & Zheludev, N. I. Nondiffracting supertoroidal pulses: optical "Kármán vortex streets". *Nat. Commun.* **15**(1), 4863 (2024).

28. Raybould, T., Fedotov, V. A., Papasimakis, N., Youngs, I., & Zheludev, N. I.. Exciting dynamic anapoles with electromagnetic doughnut pulses. *Appl. Phy. Lett.* **111**(8), 081104 (2017).

29. Wang, R., Bao, P.-Y., Hu, Z.-Q., Wang, B.-Z., & Shen, Y. Single-antenna 3D localization with nonseparable toroidal pulses, arXiv:2405.05979, 2024.

30. Wan, C., Cao, Q., Chen, J., Chong, A., & Zhan, Q. Toroidal vortices of light. *Nat. Photon.* **16**(7), 519-522 (2022).

31. Chong, A., Wan, C., Chen, J., & Zhan, Q. Generation of spatiotemporal optical vortices with controllable transverse orbital angular momentum. *Nat. Photon.* **14**(6), 350-354 (2020).

32. Zhang, H., Sun, Y., Huang, J., Wu, B., Yang, Z., Bliokh, K. Y., & Ruan, Z. Topologically crafted spatiotemporal vortices in acoustics. *Nat. Commun.* **14**(1), 6238 (2023).

33. Ge, H., Liu, S., Xu, X. Y., Long, Z. W., Tian, Y., Liu, X. P., ... & Chen, Y. F. Spatiotemporal acoustic vortex beams with transverse orbital angular momentum. *Phys. Rev. Lett.* **131**(1), 014001 (2023).

34. Che, Z., Liu, W., Ye, J., Shi, L., Chan, C. T., & Zi, J. Generation of spatiotemporal vortex pulses by resonant diffractive grating. *Phys. Rev. Lett.* **132**(4), 044001 (2024).

35. Huo, P., Chen, W., Zhang, Z., Zhang, Y., Liu, M., Lin, P., ... & Xu, T. Observation of spatiotemporal optical vortices enabled by symmetry-breaking slanted nanograting. *Nat. Commun.* **15**(1), 3055 (2024).

36. Jana, K., Mi, Y., Møller, S. H., et al. Quantum control of flying doughnut terahertz pulses. *Sci. Adv.* **10**(2), eadl1803 (2024).

37. Wang, R., Bao, P. Y., Hu, Z. Q., Shi, S., Wang, B. Z., Zheludev, N. I., & Shen, Y. Observation of resilient propagation and free-space skyrmions in toroidal electromagnetic pulses. *Appl. Phy. Rev.* **11**(3) (2024).

38. Yang, D. J., Li, Y., Zhang, Y. Q., Liu, L., Xie, Y. H., Fu, X., ... & Wang, Q. Q. Plasmonic Toroidal Vortices. *Laser Photon. Rev.* 2400474 (2024).

39. Cheng, J., Liu, W., Wu, Z., & Wan, C. Compact device for the generation of toroidal spatiotemporal optical vortices. *Opt. Lett.* **49**(16), 4646–4649 (2024).

40. Chen, W., Liu, Y., Yu, A. Z., Cao, H., Hu, W., Qiao, W., ... & Lu, Y. Q. Observation of chiral symmetry breaking in toroidal vortices of light. *Phys. Rev. Lett.* **132**(15), 153801 (2024).

41. Hellwarth, R. W. & Nouchi, P. Focused one-cycle electromagnetic pulses. *Phys. Rev. E* **54**, 889–895 (1996).

42. Xu, C., Wang, Y., Zhang, C., Dagens, B., and Zhang, X. Optical spatiotemporal differentiator using a bilayer plasmonic grating. *Opt. Lett.* **46**, 4418-4421 (2021).

43. Muelas-Hurtado, R. D., Volke-Sepúlveda, K., Ealo, J. L., et al. Observation of polarization singularities and topological textures in sound waves. *Phys. Rev. Lett.* **129**, 204301 (2022).

44. Shen, Y., Zhang, Q., Shi, P., Du, L., Zayats, A. V., & Yuan, X. Optical skyrmions and other topological quasiparticles of light. *Nat. Photon.* **18**, 15–25 (2024).

## Acknowledgements

The authors acknowledge the supports of the the National Natural Science Foundation of China (62171081, 61901086, U2341207), the Aeronautical Science Foundation of China (2023Z062080002), and the Natural Science Foundation of Sichuan Province (2022NSFSC0039). Y. Shen also acknowledges the support from Nanyang Technological University Start Up Grant, Singapore Ministry of Education (MOE) AcRF Tier 1 grant (RG157/23), and MoE AcRF Tier 1 Thematic grant (RT11/23).

## Contributions

R.W. conceived the ideas, R.W. and Y.S. supervised the project, B.Y. and R.W. performed the theoretical modeling and numerical simulations, R.W. developed the experimental methods, S.S., B.Y. and R.W. conducted the experimental measurements, R.W., B.Y. and Y.S. conducted data analysis. All authors wrote the manuscript and participated the discussions.

## Competing interests

The authors declare no competing interests.

## Data and materials availability

The data that support the findings of this study are available from the corresponding author upon reasonable request.

## Additional information

**Supplementary information** is available for this paper. Correspondence and requests for materials should be addressed to R.W. and Y.S..

**Figures**

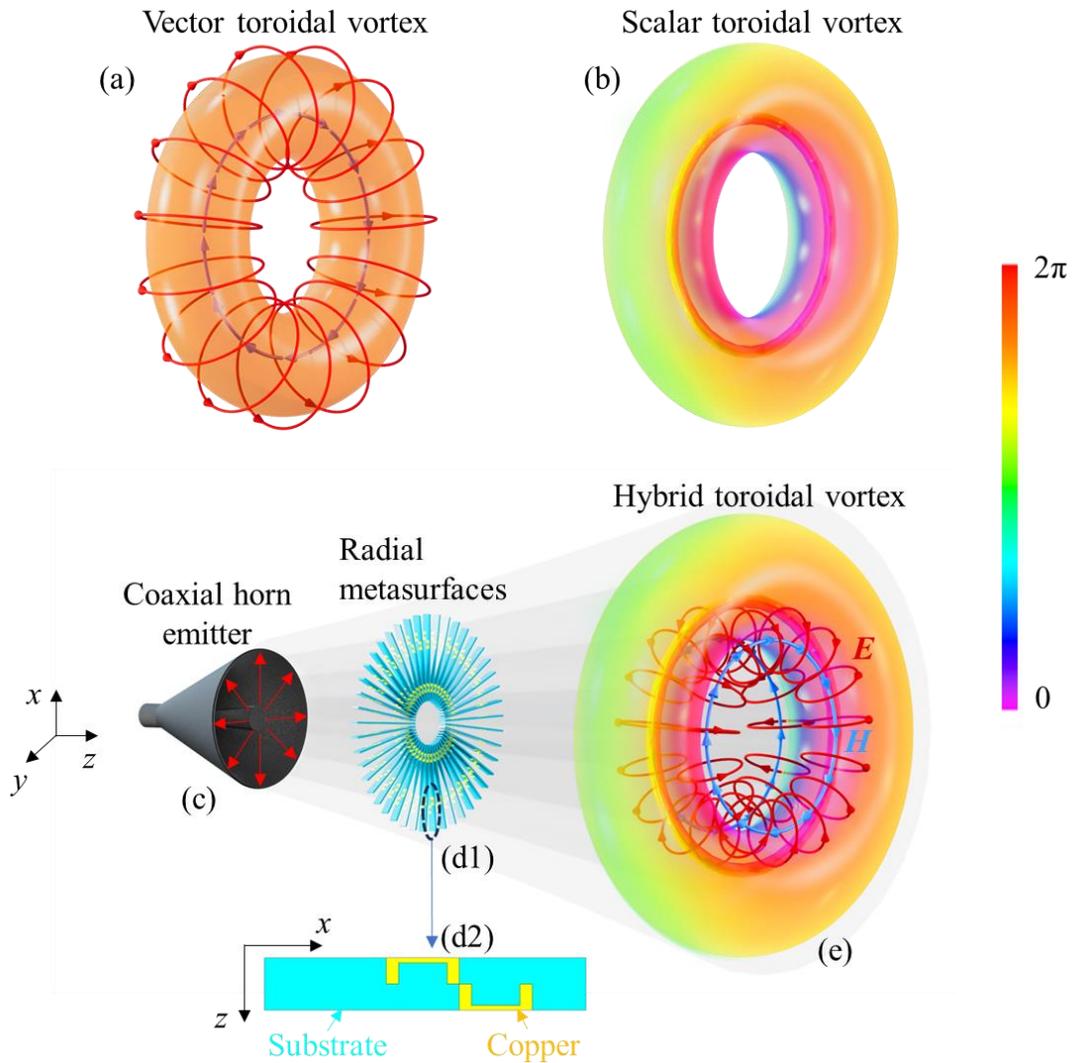

Vector toroidal vortex

(a)

Scalar toroidal vortex

(b)

Hybrid toroidal vortex

Coaxial horn emitter

Radial metasurfaces

$2\pi$

$0$

(c)

(d1)

(d2)

Substrate    Copper

**Fig. 1. Scheme for generating HETVs.** (a) Vector and (b) scalar toroidal vortices. (c) Coaxial horn emitter with inner and outer conductors, enabling the generation of radially polarized pulses. (d1) Radial metasurfaces with (d2) radially arranged units. The radial metasurface converts radially polarized waves into scalar toroidal vortices, inducing a phase variation of $2\pi$ around the surface of a spatiotemporal torus along a poloidal coordinate. The singularities of the scalar toroidal vortex induce saddle points, resulting in the generation of vector electromagnetic toroidal

vortices, delineating a toroidal surface with electric field lines. The vector and scalar toroidal vortices exhibit coupled topologies and spatial relationships, forming a new type of electromagnetic pulse: (e) HETV.

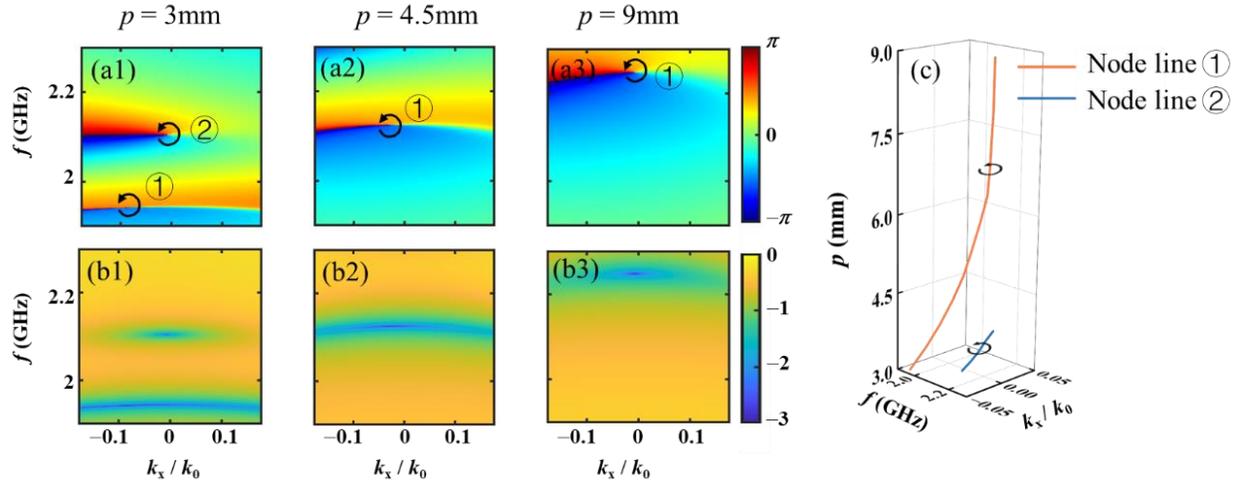

**Fig. 2 Transmission spectrum function of asymmetric metasurface unit.** (a1-a3) Phase response and (b1-b3) amplitude response. (c) Singularities for different $p$ values. When the subarray arrangement period varies between 2.4 and 7.7 (corresponding to the distance range of the metallic structures in each radially arranged subarray), there is always one vortex phase singularity with the same handedness in the 2-2.2 GHz range. This ensures that when incident pulses are in the 2-2.2 GHz range, the radial metasurfaces can convert radially polarized electromagnetic waves into spatiotemporal vortices at each radial position, thereby forming a scalar toroidal vortex.

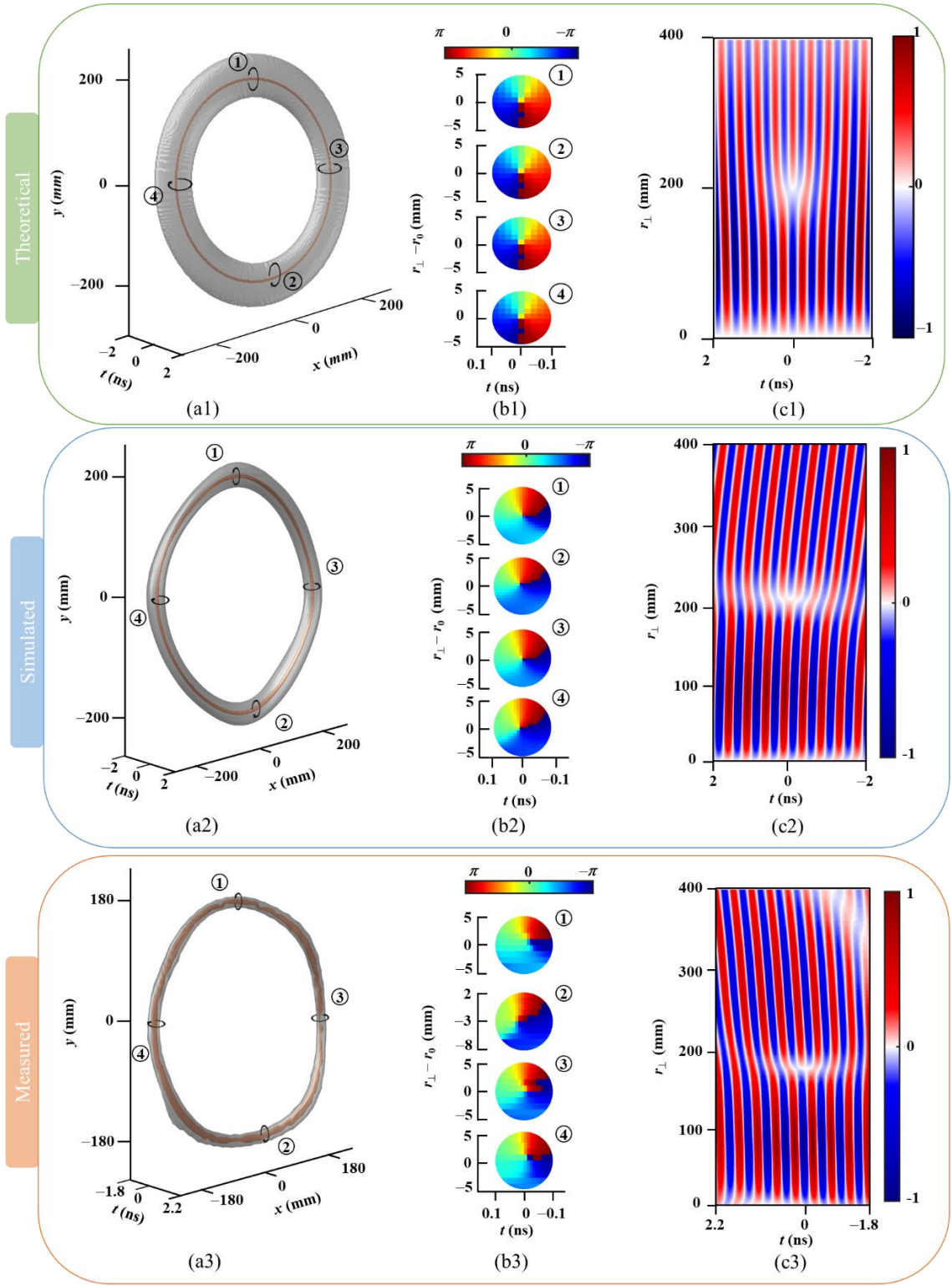

**Fig. 3 Intensity and phase characterization of HETVs.** (a1) Theoretical, (a2) simulated, and (a3)

measured 3D iso-intensity profile of the radial polarized electric field components of the toroidal vortex pulse. The outer layer of the isosurface is depicted in gray, while the vortex core surface is highlighted in brown for clarity. Theoretical, simulated, and measured phase profiles of the 4 slices in local coordinates are respectively shown in (b1), (b2), and (b3), revealing a $2\pi$ spiral phase. (c1) Theoretical, (c2) simulated, and (c3) measured 2D spatiotemporal electric field distribution at an $(r, t)$ plane, demonstrating the presence of a fork-shaped pattern indicating the vortex singularity.

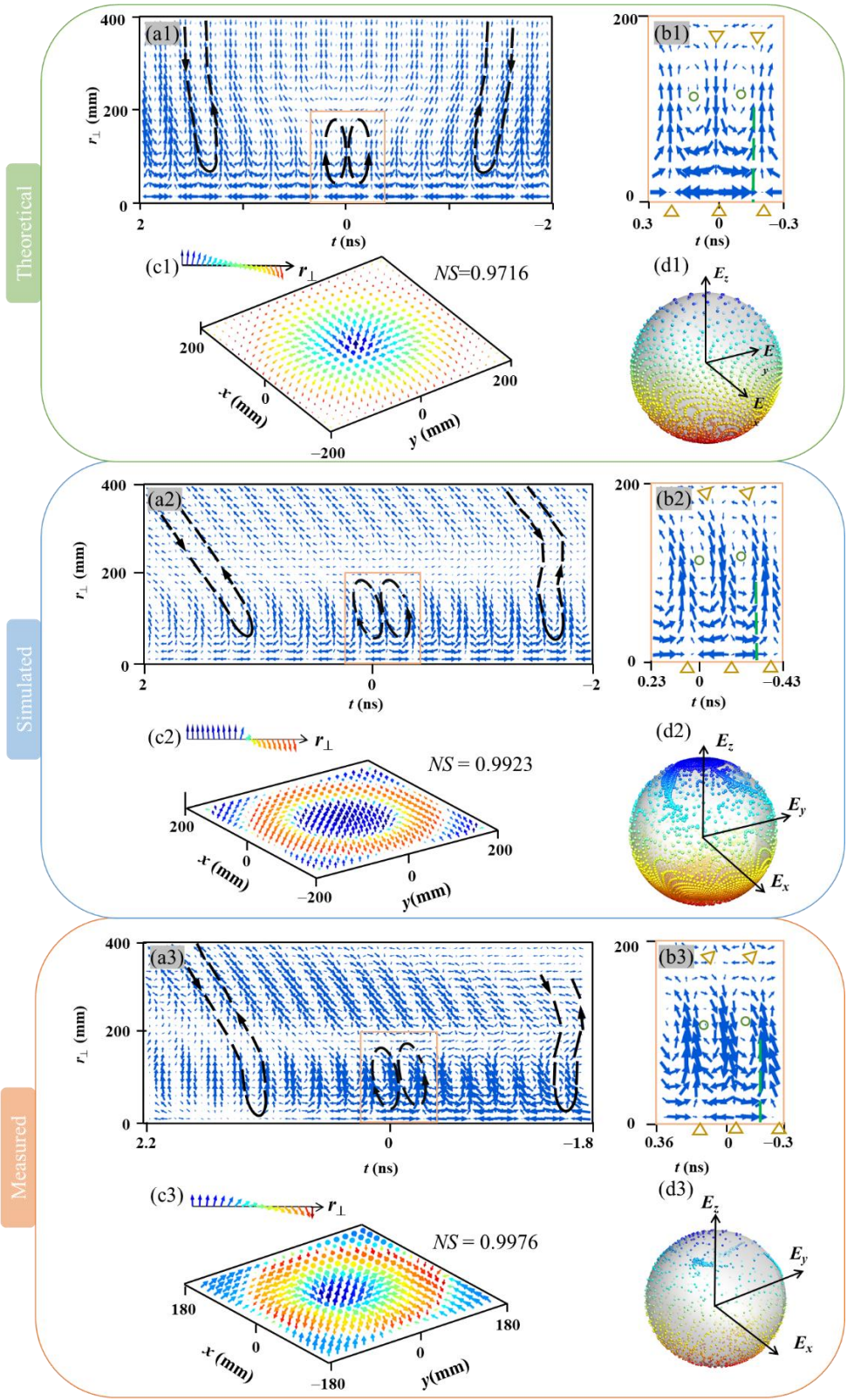

**Fig. 4 Vector field distribution of HETVs.** (a1) Theoretical, (a2) simulated, and (a3) measured

2D spatiotemporal electric field vector distribution at an $(r, t)$ plane and enlarged (b1) theoretical,

(b2) simulated, and (b3) measured portion around the vortex singularity. Similar vector field distributions can be observable at various radii, characteristic of a vector toroidal vortex. The field displays vector singularities, such as saddle points ("longitudinal-toward radial-outward" or "radial-toward longitudinal-outward", indicated by "△") and vortex rings extending away from the central axis (where electric vectors encircle to form a vortex loop, marked by "○"). The green dashed lines in (b1), (b2), and (b3) denote the positions of the theoretical, simulated, and measured skyrmionic textures in (c1), (c2), and (c3) at specific times on the $xy$ plane, respectively. The skyrmion number (NS) is approximately 1, indicating well-defined skyrmionic textures. The coverage of the sphere of field vectors in (d1), (d2), and (d3), respectively corresponding to the skyrmionic textures in (c1), (c2), and (c3), spans the surface of the sphere, confirming the presence of skyrmions.